\documentclass[%
aps,
prb,
 amsmath,amssymb,
 superscriptaddress,
preprint,%
longbibliography
]{revtex4-1}

\usepackage{graphicx}
\usepackage{dcolumn}
\usepackage{bm}
\usepackage{textcomp}
\usepackage{xcolor}
\usepackage{bm}
\usepackage[sectionbib]{bibunits} 

\newcommand{\IPI}{\affiliation{I. Physikalisches Institut, Georg-August-Universit\"at G\"ottingen, Friedrich-Hund-Platz 1, 37077 G\"ottingen, Germany}}

\newcommand{\graz}{\affiliation{Institute of Physics, University of Graz, NAWI Graz, Universitätsplatz 5, 8010 Graz, Austria}}
\newcommand{\KL}{\affiliation{Department of Physics and Research Center OPTIMAS, University of Kaiserslautern,
Erwin-Schr\"odinger-Straße 46, 67663, Kaiserslautern, Germany}}
\newcommand{\ICASEC}{\affiliation{International Center for Advanced Studies of Energy Conversion (ICASEC), University of Göttingen, Göttingen, Germany}}
\newcommand{\CNRS}{\affiliation{Univ. Grenoble Alpes, CNRS, Inst NEEL, F-38042 Grenoble, France}}
\newcommand{\mainz}{\affiliation{Institute of Physics, Johannes Gutenberg-University Mainz, 55128 Mainz, Germany}}

\begin{document}
\newcommand{\csixty}{C\textsubscript{60}}

\newcommand{\sone}{S\textsubscript{1}}
\newcommand{\stwo}{S\textsubscript{4}}
\newcommand{\ctone}{S\textsubscript{2}}
\newcommand{\cttwo}{S\textsubscript{3}}

\newcommand{\ekin}{$E_\mathrm{kin}$}
\newcommand{\ppduration}{40~fs}
\newcommand{\invA}{\AA\textsuperscript{-1}}

\newcommand{\ve}[1]{\boldsymbol{#1}} 
\newcommand{\ud}{\mathrm{d}}


\title{
Multiorbital exciton formation in an organic semiconductor
}

\author{Wiebke Bennecke} %
\IPI

\author{Andreas Windischbacher}
\graz

\author{David Schmitt} %
\author{Jan Philipp Bange} %
\IPI
\author{Ralf Hemm} %
\KL
\author{Christian~S. Kern}
\graz

\author{Gabriele D'Avino}
\author{Xavier Blase}
\CNRS

\author{Daniel Steil} %
\author{Sabine Steil} 
\IPI

\author{Martin~Aeschlimann}
\KL

\author{Benjamin Stadtm\"uller} 
\KL
\mainz

\author{Marcel Reutzel}%
\IPI

\author{Peter Puschnig}
\graz

\author{G.~S.~Matthijs~Jansen} \email{gsmjansen@uni-goettingen.de} %
\IPI

\author{Stefan Mathias}

\email{smathias@uni-goettingen.de}%
\IPI
\ICASEC

\date{\today}

\begin{abstract} 
Harnessing the optoelectronic response of organic semiconductors requires a thorough understanding of the fundamental light-matter interaction that is dominated by the excitation of correlated electron-hole pairs, i.e. excitons. The nature of these excitons would be fully captured by knowing the quantum-mechanical wavefunction, which, however, is difficult to access both theoretically and experimentally. Here, we use femtosecond photoemission orbital tomography in combination with many-body perturbation theory to gain access to exciton wavefunctions in organic semiconductors. 
We find that the coherent sum of multiple electron-hole pair contributions that typically make up a single exciton can be experimentally evidenced by photoelectron spectroscopy. For the prototypical organic semiconductor buckminsterfullerene (\csixty{}), we show how to disentangle such multiorbital contributions and thereby access key properties of the exciton wavefunctions including localization, charge-transfer character, and ultrafast exciton formation and relaxation dynamics.
\end{abstract}

\maketitle

\section{MAIN} 
Excitons, quasiparticles consisting of bound electron-hole pairs, are at the heart of the optoelectronic response of all organic semiconductors, and exciton formation and relaxation processes are largely responsible for energy conversion and light harvesting applications in these materials. At the atomic level, excitons are described by a two-particle correlated quantum-mechanical wavefunction that includes both the excited electron and the remaining hole. This wavefunction covers the complete shape of the exciton wave and thus provides access to a number of critical exciton properties such as the orbital character, the degree of (de)localization, the degree of charge separation, and whether this involves charge transfer between molecules. Consequently, in order to fully understand exciton dynamics and to exploit them in, e.g., an organic solar cell, an accurate and complete measurement of the exciton wavefunction would be ideal. Exemplary in this situation is the ongoing work to understand the optoelectronic response of \csixty{}, a prototypical organic semiconductor that is commonly used in organic solar cells \cite{ke_efficient_2015, puente_santiago_tailoring_2020, yu_simplified_2020}. Here, a topic of research has been the optical absorption feature that occurs at 2.8~eV for multilayer and other aggregated structures of \csixty{} \cite{wang_aggregates_1993}. Interestingly, time- and angle-resolved photoelectron spectroscopy and optical absorption spectroscopy studies have indirectly found that this optical transition corresponds to the formation of charge-transfer excitons with significant electron-hole separation \cite{stadtmuller_strong_2019, emmerich_ultrafast_2020, hess_electroabsorption_1996, causa_femtosecond_2018, hahn_role_2016}. Although these hints are supported by time-dependent density functional theory calculations that show the importance of delocalized excitations in \csixty{} clusters \cite{mumthazmuhammed_impact_2022, habeebmokkath_delocalized_2021, kobayashi_wannier-like_2020}, quantitative measurements of the exciton localization and charge separation have so far not been possible. Thus, the \csixty{} case highlights the need for a more direct experimental access to the wavefunctions of the electron-hole pair excitations.

From an experimental point of view, our method of choice to access exciton wavefunctions is time-resolved photoemission orbital tomography (tr-POT, see Methods for the experimental realization used in this work)\cite{puschnig_reconstruction_2009, jansen_efficient_2020, wallauer_tracing_2021, neef_orbital-resolved_2022}. In POT, the comparison with density functional theory calculations (DFT) provides a direct connection between photoemission data and the orbitals of the electrons.\cite{puschnig_reconstruction_2009} The extension to the time-domain promises valuable access also to the spatial information of excited electrons. However, at least for organic semiconductors, it has not been explicitly considered that photoemission of excitons requires the break-up of the two-particle electron-hole pair and that only the photoemitted electron, but not the hole, is directly detected. In fact, it is not clear to what extent (tr-)POT can be reasonably used for the interpretation and analysis of such strongly interacting correlated quasiparticles. Here we address this open question and show how tr-POT can probe the exciton wavefunction in the example system of a \csixty{} multilayer.

\begin{figure*}[tb!]
    \centering
    \includegraphics[width=.7\textwidth]{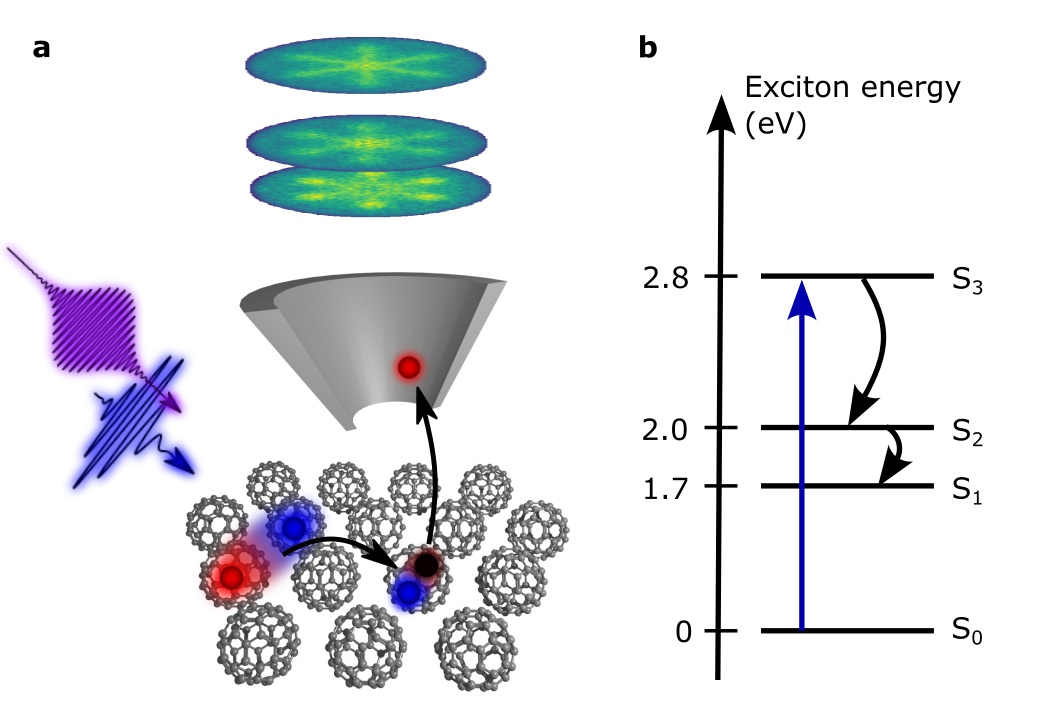}
    \caption{
    Schematic overview of time-resolved photoemission orbital tomography of exciton states in \csixty{}. \textbf{a, b}, a femtosecond optical pulse (blue pulse and blue arrow in (\textbf{a}) and (\textbf{b}), respectively) excites optically bright excitons in a \csixty{} film. The exciton electron-hole pairs are sketched in (\textbf{a}) as correlated particles (shaded blue-red areas) with a blue sphere for the hole and a red sphere for the electron. We probe the excitons in \csixty{} with extreme ultraviolet (EUV) pulses (purple pulse in (\textbf{a}), $h\nu =$~26.5~eV) that break up the electron-hole pairs and photoemit the corresponding electrons (red spheres), of which we detect the kinetic energies and the momentum emission patterns (yellow-green-colored disks in (\textbf{a})).
    The optical excitation of excitons in \csixty{} is known to lead to the formation of a decay sequence of singlet exciton states with varying charge-transfer character\cite{stadtmuller_strong_2019, emmerich_ultrafast_2020} (see (\textbf{b}), S\textsubscript{i}: i\textsuperscript{th} singlet excited state, S\textsubscript{0}: ground state). We are able to measure these exciton dynamics and the corresponding orbital tomography momentum patterns by adjusting the temporal delay between the optical excitation and the EUV probe pulses.
    }
    \label{fig:exp_excitons}
\end{figure*}

We employ our recently developed setup for photoelectron momentum microscopy\cite{medjanik_direct_2017, keunecke_time-resolved_2020, Keunecke20prb} and use ultrashort laser pulses to optically excite bright excitons in \csixty{} thin films that were deposited on Cu(111) (measurement temperature T $\approx$ 80~K; see Methods and Figure~1a). 
In the time-resolved photoemission experiment, we detect the energy and momentum emission pattern of the photoemitted electrons, which were initially part of the bound electron-hole pairs, i.e. the excitons. Following the time-evolution of the photoelectron spectrum, we can observe how the optically excited states relax to energetically lower-lying dark exciton states with different localization and charge-transfer character\cite{stadtmuller_strong_2019, emmerich_ultrafast_2020} (Figure~1b and data in Extended Fig.~\ref{fig:expdynamics}). 
In addition to the energy relaxation, we collect tr-POT data, and investigate in how far these patterns can be used to access real-space properties of the exciton wavefunctions. Specifically, we will address two questions: which orbitals contribute to the formation of the excitons and how this key information is imprinted in the energy- and momentum-resolved photoemission spectra.

\section{Results \& Discussion}
\subsection{The exciton spectrum of buckminsterfullerene \csixty{}}
\begin{figure*}[tbh!]
    \centering
    \includegraphics[width=\linewidth]{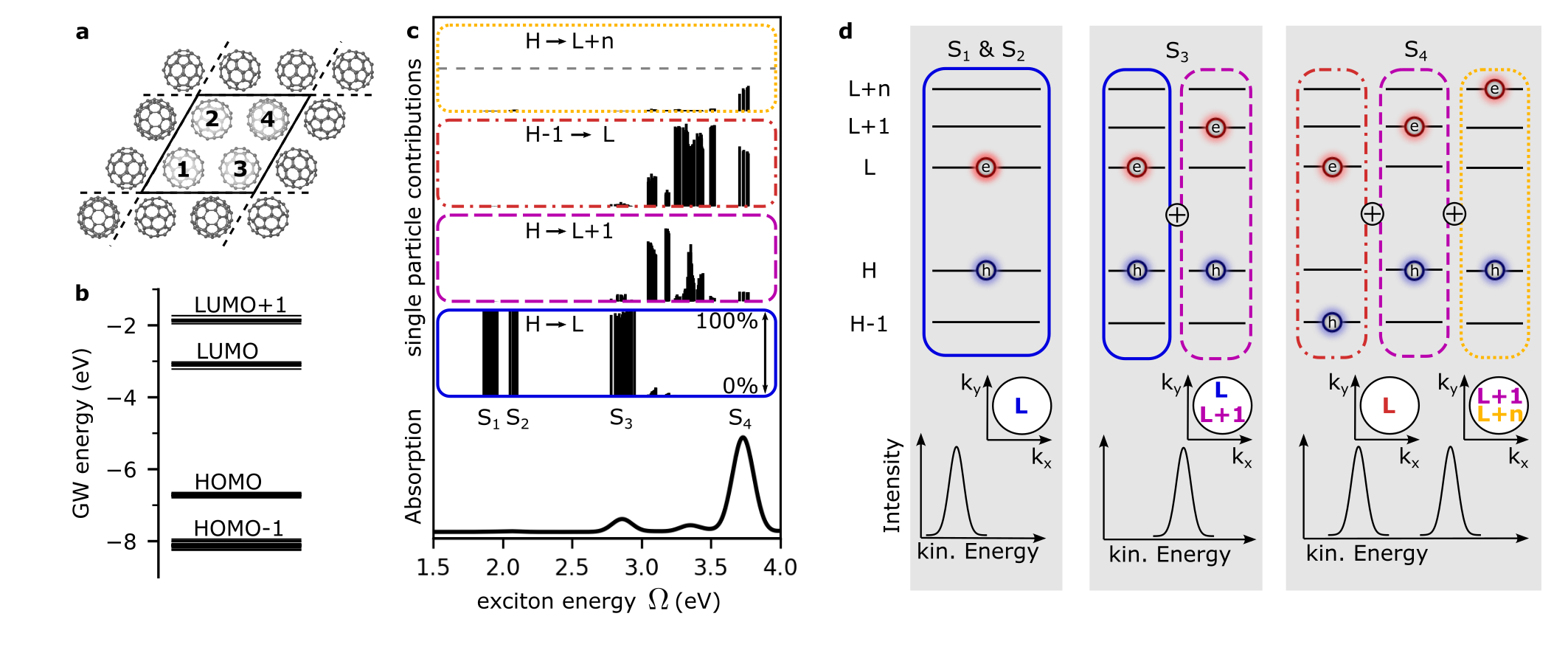}
    \caption{\textit{Ab-initio} calculation of the electronic structure and exciton spectrum of \csixty{} dimers in a crystalline multilayer sample. \textbf{a}, the unit cell for a monolayer of \csixty{}, for which $GW$+BSE calculations for the dimers 1-2 and 1-4 were performed. \textbf{b}, electron addition/removal single-particle energies as retrieved from the self-consistent $GW$ calculation. These energies directly provide $\varepsilon_v$ in Eq.~\eqref{eq:trPOT}. \textbf{c}, results of the full $GW$+BSE calculation, showing as a function of the exciton energy $\Omega$ from bottom to top: the calculated optical absorption, the exciton band assignment \sone{} - \stwo{}, and the relative contributions to the exciton wavefunctions of different electron-hole pair excitations $\phi_v(\ve{r}_h) \chi_c(\ve{r}_e)$. Full details on the calculations are given in the Methods section. \textbf{d}, sketch of the composition of the exciton wavefunction of the \sone{} - \stwo{} bands and their expected photoemission signatures based on Eq.~\eqref{eq:trPOT}. In order to visualize the contributing orbitals, blue holes and red electrons are assigned to the single-particle states as shown in (\textbf{b}).}
    \label{fig:bse_excitons}
\end{figure*}

To lay the foundation for our study, we first discuss the theoretical electronic properties of the \csixty{} film. On top of a hybrid-functional DFT ground state calculation, we obtain the exciton spectrum by employing the many-body framework of $GW$ and Bethe-Salpeter-Equation ($GW$+BSE) calculations (see Methods for full details). As shown in Fig.~\ref{fig:bse_excitons}a, we model the \csixty{} low-temperature phase by the two symmetry-inequivalent \csixty{} dimers 1-2 and 1-4, respectively\cite{wang_orientational_2001}, which are properly embedded to account for polarization effects in the film. The calculated single-particle energy levels are shown in Fig.~\ref{fig:bse_excitons}b, where we group the electron removal and electron addition energies into four bands, denoted as HOMO-1, HOMO, LUMO, and LUMO+1 according to the orbitals of the parent orbitals of an gas-phase \csixty{} molecule. These manifolds consist of 18, 10, 6, and 6 energy levels per dimer, respectively, originating from the  $g_g$+$h_g$, $h_u$, $t_{1u}$, and $t_{1g}$ irreducible representations of the gas phase \csixty{} orbitals \cite{dresselhaus_c60_1996}. We emphasize that the calculated $GW$ ionization levels of HOMO and HOMO-1 of 6.7~eV and 8.1~eV are in excellent agreement with experimental data for this \csixty{} film (see SI and Ref.~\onlinecite{haag_signatures_2020}).

Building upon the $GW$ single-particle energies, we solve the Bethe-Salpeter equation and compute the energies $\Omega_m$ of all correlated electron-hole pairs (excitons). The resulting absorption spectrum (bottom panel of Fig. \ref{fig:bse_excitons}c) agrees well with literature\cite{wang_aggregates_1993}. In addition, we obtain the weights $X^{(m)}_{v c}$ on the specific electron-hole pairs that coherently contribute to the $m$\textsuperscript{th} exciton state, from which the exciton wavefunctions are constructed in the Tamm-Dancoff approximation as follows:
\begin{equation}
\psi_m(\ve{r}_h, \ve{r}_e) = \sum_{v,c} X^{(m)}_{vc} \phi_v^*(\ve{r}_h) \chi_c(\ve{r}_e).
\label{eq:exciton_state}
\end{equation}
This means that each exciton $\psi_m$ with energy $\Omega_m$ consists of a weighted coherent sum of multiple electron-hole-transitions $\phi_v(\ve{r}_h) \chi_c(\ve{r}_e)$ each containing one electron orbital $\chi_c$ and one hole orbital $\phi_v$.

To gain more insight into the character of the excitons $\psi_m$, we qualitatively classify them according to the most dominant orbital contributions that are involved in the transitions. This is visualized in the four sub-panels above the absorption spectrum in Fig.~\ref{fig:bse_excitons}c. For a given exciton energy $\Omega_m$, the black bars in each sub-panel show the partial contribution $|X_{v,c}^{(m)}|^2$ of characteristic electron-hole transitions $\phi_v(\ve{r}_h) \chi_c(\ve{r}_e)$ to a given exciton $\psi_m$. Looking at individual sub-panels, we see first that characteristic electron-hole transitions can belong to different excitons $\psi_m$ that have very different exciton energies $\Omega_m$. For example, the blue panel in Fig.~2c shows the contributions of HOMO$\rightarrow$LUMO (abbreviated H$\rightarrow$L) transitions as a function of exciton energy $\Omega_m$, and we see that these transitions contribute to excitons that are spread in energy over a scale of more than 1~eV (from $\Omega_m$ $\approx$ 1.7~eV to 3~eV). This spread of H$\rightarrow$L contributions (and also H$-$n$\rightarrow$L+m contributions) is caused by the fact that there are already many orbital energies per dimer (cf. Fig.~2b) which combine to form excitons with different degrees of localization and delocalization of the electrons and holes on one or more molecules.

We now focus on four exciton bands of the \csixty{} film, denoted as \sone{} - \stwo{}, which are centered around $\Omega_\sone{}$, $\Omega_\ctone{}$, $\Omega_\cttwo{}$ and $\Omega_\stwo{}$ at 1.9, 2.1, 2.8 and 3.6~eV, respectively (cf. Ref.~\onlinecite{emmerich_ultrafast_2020}). 
It is important to emphasize that each exciton band \sone{} - \stwo{} arises from many individual excitons $\psi_m$ with similar exciton energies $\Omega_m$ within the exciton band. Looking again at the sub-panels, we see that the \sone{} and \ctone{} exciton bands are made up of excitons $\psi_m$ that are almost exclusively composed of transitions from H$\rightarrow$L. On the other hand, the \cttwo{} shows in addition to H$\rightarrow$L also significant contributions from H$\rightarrow$L+1 transitions (pink-dashed panel). The \stwo{} exciton band can be characterized as arising from H$\rightarrow$L+1 (pink-dashed panel) and H$-$1$\rightarrow$L (orange-dash-dotted panel) as well as transitions from the HOMO to several higher lying orbitals denoted as H$\rightarrow$L+n (yellow-dotted panel). We emphasize that although orbitals from several different $GW$ energies contribute, e.g., to an exciton in the \stwo{} band, the exciton energy $\Omega_m$ of each exciton $\psi_m$ has a single well-defined value.

\subsection{Photoemission signatures of multiorbital contributions}
In the following, we investigate whether these theoretically predicted multiorbital characteristics of the excitons can also be probed experimentally. As will be shown below, time-resolved photoemission spectroscopy can indeed provide access not only to the dark exciton landscape \cite{weinelt_dynamics_2004, dong_direct_2021, wallauer_momentum-resolved_2021, Madeo20sci, schmitt_formation_2022}, but also to the distinct orbital contributions of exciton states. A key step in extracting this information from the experimental data lies in a thorough comparison with simulations that specifically consider the pump-probe photoemission process, a topic that has recently attracted increased attention \cite{Popova2016, DeGiovannini2017, hammon_pump-probe_2021, Reuner2023}. Here, we rely on the formalism of Kern \textit{et al.}\cite{kern_exciton_tomog}, which is based on a common Fermi’s golden rule approach to photoemission \cite{Dauth2014}. Assuming the exciton of Eq.~\eqref{eq:exciton_state} as the initial state and applying the plane-wave final state approximation of POT, the photoemission intensity of the exciton $\psi_m$ is formulated as
\begin{equation}
    I_m(E_\mathrm{kin},\ve{k})  \propto  
    \left| \ve{A} \ve{k} \right|^2
    \sum_{v} \left| \sum_{c} X^{(m)}_{v c}  \mathcal{F}\left[\chi_{c} \right] (\ve k)\right|^2 
     \times   \delta\left(h\nu - E_\mathrm{kin} - \varepsilon_v + \Omega_m \right).
\label{eq:trPOT}
\end{equation}
Here $\ve{A}$ is the vector potential of the incident light field, $\mathcal{F}$ the Fourier transform, $\ve{k}$ the photoelectron momentum, $h\nu$ the probe photon energy, $\varepsilon_v$ the $v$\textsuperscript{th} ionization potential, $\Omega_m$ the exciton energy, and $E_{kin}$ the energy of the photoemitted electron. Note that $\varepsilon_v$ directly indicates the final-state energy of the left-behind hole. In the context of our present study, delving into Eq.~\eqref{eq:trPOT} leads to two striking consequences, which we discuss in the following. 

First, we illustrate the consequences of the multiorbital character of the exciton states on the photoelectron spectrum, and sketch in Fig.~2d the typical single-particle energy level diagrams for the HOMO and LUMO states and then indicate the contributing orbitals to the two-particle exciton state by blue holes and red electrons in these states, respectively. For the \sone{} exciton band (left panel), we already found that the main orbital contributions to the band are of H$\rightarrow$L character (Fig. 2d, left, and cf. Fig. 2c, blue panel). To determine the kinetic energy of the photoelectrons originating from the exciton, we have to consider the correlated nature of the electron-hole pair. The energy conservation expressed by the delta function in Eq.~2 (see also Ref.~\onlinecite{weinelt_dynamics_2004, Zhu14jpcl}) requires that the kinetic energy of the photoelectron depends on the ionization energy of the involved HOMO hole state $\varepsilon_{v}$ = $\varepsilon_{H}$ and the correlated electron-hole pair energy $\Omega \approx \Omega_{S_{1}}$. Therefore, we expect to measure a single photoelectron peak, as shown in the lower part of the left panel of Fig.~2d. In the case of the \ctone{} exciton the situation is similar, since the main orbital contributions are also of H$\rightarrow$L character. However, since the \ctone{} exciton band has a different energy $\Omega_{\ctone{}}$, the photoelectron peak is located at a different kinetic energy with respect to the S$_1$ peak.  

In the case of the \cttwo{} exciton band, we find that in contrast to the \sone{} and \ctone{} excitons not only H$\rightarrow$L, but also H$\rightarrow$L+1 transitions contribute (Fig. 2d, middle panel, and cf. Fig~2c, blue and pink-dashed panels, respectively). However, we still expect a single peak in the photoemission, because the same hole states are involved for both transitions (i.e., same $\varepsilon_v = \varepsilon_{H}$ in the sum in Eq.~2), and all orbital contributions have the same exciton energy $\approx\Omega_{\cttwo{}}$, even though transitions with electrons in energetically very different single-particle LUMO and LUMO+1 states contribute. With other words, and somewhat counter-intuitively, the single-particle energies of the electron orbitals (the LUMOs) contributing to the exciton do not enter the energy conservation term in Eq.~2, and thus do not affect the kinetic energy observed in the experiment. 

Finally, for the \stwo{} exciton band at $\Omega_\stwo{} =$~3.6~eV, we find three major contributions (Fig.~2d, right panel), where not only the electrons but also the holes are distributed over two energetically different levels, namely the HOMO (cf. pink-dashed and yellow-dotted panels in Fig.~2c,d) and the HOMO-1 (cf. orange-dash-dotted panels in Fig.~2c,d). Thus, there are two different final states available for the hole, each with a different binding energy.
Consequently, the photoemission spectrum of \stwo{} is expected to exhibit a double-peak structure with intensity appearing $\approx$~3.6~eV above the HOMO kinetic energy $E_{H}$, and $\approx$~3.6~eV above the HOMO-1 kinetic energy $E_{H-1}$, as illustrated in the right-most panel of Fig.~\ref{fig:bse_excitons}d. Relating this specifically to the single-particle picture of our $GW$ calculations, the two peaks are predicted to have a separation of $\varepsilon_{{H-1}}$ – $\varepsilon_{{H}}$ = 8.1 – 6.7 = 1.4~eV. 

In addition, Eq.~\eqref{eq:trPOT} now also provides the theoretical framework for interpreting momentum-resolved tr-POT data from excitons. Ground state POT can be easily understood in terms of the Fourier transform $\mathcal{F}$ of single-particle orbitals. A naive extension to excitons might imply an incoherent, weighted sum of all LUMO orbitals $\chi_c$ contributing to the exciton wavefunction. However, as Eq.~\eqref{eq:trPOT} shows, such a simple picture proves insufficient. Instead, the momentum pattern of the exciton wavefunction is related to a coherent superposition of the electron orbitals $\chi_c$ weighted by the electron-hole coupling coefficients $X^{(m)}_{v c}$. The implications of this finding are sketched in the $k_x$-$k_y$ plots in Fig.~\ref{fig:bse_excitons}d and are most obvious for the \cttwo{} band. Here, the exciton is composed of transitions with a common hole position, i.e., H$\rightarrow$L and H$\rightarrow$L+1, leading to a coherent superposition of all 12 electron orbitals from the LUMO and LUMO+1 in the momentum distribution. In summary, multiple hole contributions can be identified in a multi-peak structure in the photoemission spectrum, and multiple electron contributions will result in a coherent sum of the electron orbitals that can be identified in the corresponding energy-momentum patterns from tr-POT data.

\subsection{Disentangling multiorbital contributions experimentally}
\begin{figure*}[tbh!]
    \centering
    \includegraphics[width=1.\textwidth]{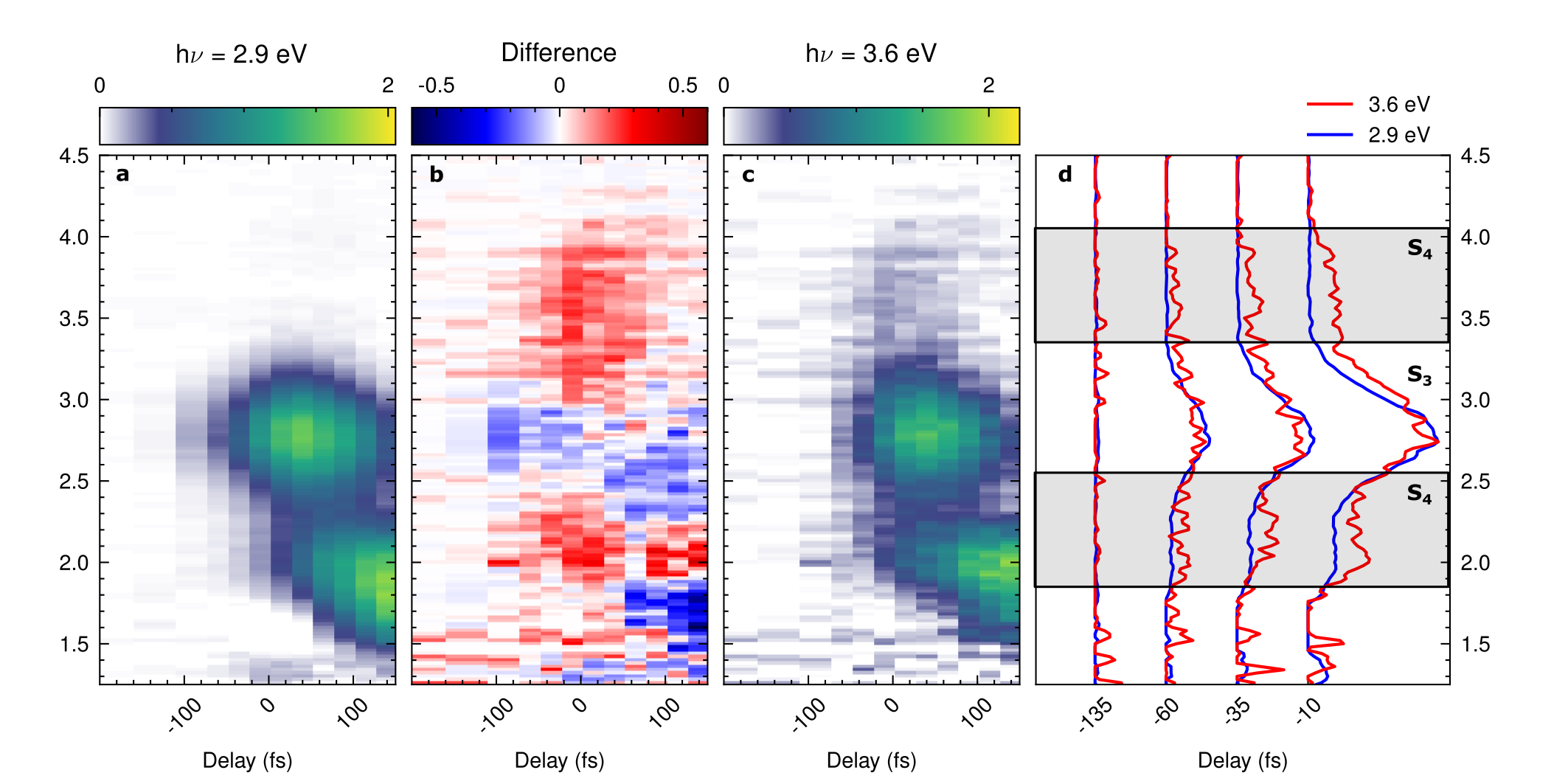}
    \caption{\textbf{a-d}, comparison of the time-resolved photoelectron spectra of multilayer \csixty{} for $h\nu =$~2.9~eV excitation and $h\nu =$~3.6~eV excitation ((\textbf{a}) and (\textbf{c}), respectively), both normalized and shifted in time to match the intensity of the \cttwo{} signals (see Methods for full details on the data analysis). As can be seen in the difference (\textbf{b}), for $h\nu =$~3.6~eV pump we observe an enhancement of the photoemission yield around $E-E_H \approx$~3.6~eV as well as around $E-E_H \approx$~2.2~eV. We attribute this signal to the \stwo{} exciton band, which has hole contributions stemming from both the HOMO and the HOMO-1. To further quantify the signal of the \stwo{} exciton, (\textbf{d}) shows energy distribution curves for both measurements at early delays, showing the enhancement in the $h\nu =$~3.6~eV measurement. }
    \label{fig:exp_S2}
\end{figure*}

These very strong predictions about multi-peaked photoemission spectra due to multiorbital excitons can be directly verified in an experiment on \csixty{} by comparing spectra for resonant excitation of either the \cttwo{} or the \stwo{} excitons (cf. Fig.~\ref{fig:bse_excitons}). The corresponding experimental data are shown in Fig.~\ref{fig:exp_S2}a and \ref{fig:exp_S2}c, respectively.  
Starting from the excitation of the \cttwo{} exciton band with $h\nu =$~2.9~eV photon energy (which is sufficiently resonant to excite the manifold of exciton states that make up the \cttwo{} band around $\Omega_{\cttwo{}} =$~2.8~eV), we can clearly identify the direct excitation (at 0~fs delay) of the exciton \cttwo{} feature at an energy of E~$\approx$~2.8~eV above the kinetic energy $E_H$ of the HOMO level. Shortly after the excitation, additional photoemission intensity builds up at $E-E_H \approx$~2.0~eV and $\approx$~1.7~eV, which is known to be caused by relaxation to the \ctone{} and \sone{} dark exciton states\cite{emmerich_ultrafast_2020} and is in good agreement with the theoretically predicted energies of $E-E_H$ $\approx$~2.1~eV and $\approx$1.9~eV.

Changing now the pump photon energy to $h\nu =$~3.6~eV for direct excitation of the \stwo{} exciton band (Fig.~\ref{fig:exp_S2}c), two distinct peaks at $\approx$ 3.6~eV above the HOMO and the HOMO-1 are expected from theory. While photoemission intensity at $E-E_H \approx$~3.6~eV above the HOMO level is readily visible in Fig.~\ref{fig:exp_S2}c, the second feature at 3.6~eV above the \mbox{HOMO-1} is expected at $E-E_H \approx$~2.2~eV above the HOMO level (corresponding to $E-E_{H-1}$ $\approx$~3.6~eV) and thus almost degenerate with the aforementioned \ctone{} dark exciton band at about $E-E_H \approx$~2.0~eV, which appears after the optical excitation due to relaxation processes. Therefore, we need to pinpoint this second H$-$1$\rightarrow$L contribution to the \stwo{} exciton at the earliest time of the excitation. Indeed, a closer look around 0~fs delay shows additional photoemission intensity at about $E-E_H \approx$~2.2~eV. 
Using difference maps (Fig.~\ref{fig:exp_S2}b) and direct comparisons of energy-distribution-curves at selected time-steps (Fig.~\ref{fig:exp_S2}d), we clearly find a double-peak structure corresponding to the energy difference of $\approx$1.4~eV of the HOMO and \mbox{HOMO-1} levels. Thereby, we have shown that photoelectron spectroscopy, in contrast to other techniques (e.g., absorption spectroscopy), is indeed able to disentangle different orbital contributions of the excitons. In this way, we have validated the theoretically predicted multi-peak structure of the multiorbital exciton state that is implied by Eq.~\eqref{eq:trPOT}. We also see that the photoelectron energies in the spectrum turn out to be sensitive probes of the corresponding hole contributions of the correlated exciton states.

We note that the signature of the \cttwo{} excitons, even if not directly excited with the light pulse in this measurement, is still visible and moreover with significantly higher intensity than the multiorbital signals of the resonantly excited \stwo{} exciton band. This observation strongly suggests that there is a very fast relaxation from the \stwo{} exciton to the \cttwo{} exciton, with relaxation times well below 50~fs (see Extended Fig.~\ref{fig:expdynamics}).

\subsection{Time-resolved photoemission orbital tomography of exciton wavefunctions}
\begin{figure*}[htb!]
    \centering
    \includegraphics[width=.85\linewidth]{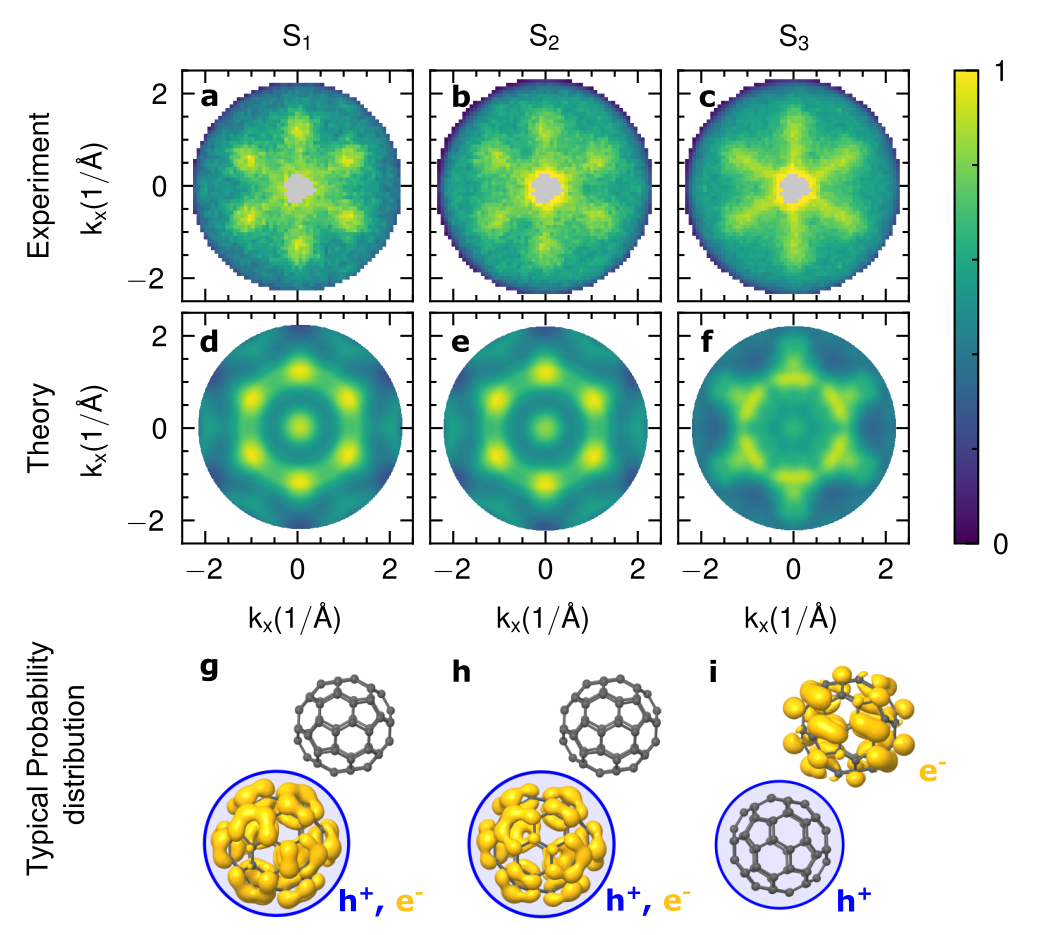}
    \caption{\textbf{a-f}, comparison of the (\textbf{a-c}) experimental momentum maps acquired for the three exciton bands observed in \csixty{} with the (\textbf{d-f}) predicted momentum maps retrieved from $GW$+BSE.
    Note that the center of the experimental maps could not be analyzed due to a space-charge-induced background signal in this region (gray area, see Methods).
    \textbf{g-i}, isosurfaces of the integrated electron probability density (yellow) within the 1-2 dimer for fixed hole positions on the bottom-left molecule (blue circle) of the dimer for the (\textbf{g}) \sone{}, the (\textbf{h}) \ctone{}, and the (\textbf{i}) \cttwo{} exciton bands. }
    \label{fig:comparison}
\end{figure*}

Based on the excellent agreement between the experiment and the $GW$+BSE theoretical results, we are now ready to investigate to what extent the momentum patterns from tr-POT data of excitons in organic semiconductors contain information about the real-space spatial distribution of the exciton wavefunction. In the experiment, we once again excite the \cttwo{} exciton band in the \csixty{} film with $h\nu =$~2.9~eV pump energy, and we now use femtosecond tr-POT to collect the momentum fingerprints of the directly excited \cttwo{} excitons around 0~fs and the subsequently built-up dark \ctone{} and \sone{} excitons that appear in the exciton relaxation cascade in the \csixty{} film (see Fig.~\ref{fig:comparison}a-c, where the momentum map of the lowest energy \sone{} exciton band is plotted in (a), the \ctone{} in (b) and the highest energy \cttwo{} exciton band in (c); see Extended Fig.~\ref{fig:expdynamics} for time-resolved traces of the exciton formation and relaxation dynamics). We note that the collection of these data required integration times of up to 70 hours, and that a measurement of the comparatively low-intensity \stwo{} feature when excited with $h\nu =$~3.6~eV has not yet proved feasible. For the interpretation of the collected POT momentum maps from the \sone{}, \ctone{}, and \cttwo{} excitons, we also calculate the expected momentum fingerprints for the wavefunctions obtained from the $GW$+BSE calculation for both dimers, each rotated to all occurring orientations in the crystal. Finally, for the theoretical momentum maps, we sum up the photoelectron intensities of each electron-hole transition in an energy range of 200~meV centered on the exciton band. The results are shown in Fig.~\ref{fig:comparison}d-f below the experimental data for direct comparison.

First, we observe that the experimental momentum maps of the \sone{} and \ctone{} states are largely similar (Fig.~\ref{fig:comparison}a,b), showing six lobes centered at $k_\parallel \approx$~1.2~\invA{}. These features, as well as the energy splitting between \sone{} and \ctone{} (cf. Fig.~\ref{fig:bse_excitons}c), are accurately reproduced by the $GW$+BSE prediction (Fig.~\ref{fig:comparison}d,e). Furthermore, also the $GW$+BSE calculation shows very similar momentum maps for \sone{} and \ctone{}, suggesting a similar spatial structure of the excitons. This is in contrast to a naive application of static POT to the unoccupied orbitals of the DFT ground state of \csixty{}, which does show a similar momentum map for the LUMO, but cannot explain a kinetic energy difference in the photoemission signal, nor give any indication of differences in the corresponding exciton wavefunctions. 
With this agreement between experiment and theory, we now extract the spatial properties of the $GW$+BSE exciton wavefunctions. To visualize the degree of charge-transfer of these two-particle exciton wavefunctions $\psi_m(\ve{r_h},\ve{r_e})$, we integrate the electron probability density over all possible hole positions $\ve{r_h}$, considering only hole positions at one of the \csixty{} molecules in the dimer. This effectively fixes the hole contribution to a particular \csixty{} molecule (blue circles in Fig.~\ref{fig:comparison}g-i indicate the boundary of considered hole positions around one molecule, hole distribution not shown), and provides a probability density for the electronic part of the exciton wavefunction in the dimer, which we visualize by a yellow isosurface (see Fig.~\ref{fig:comparison}g-i). 
Obviously, in the case of \sone{} and \ctone{} (Fig.~\ref{fig:comparison}g,h), when the hole position is restricted to one molecule of the dimer, the electronic part of the exciton wavefunction is localized at the same molecule of the dimer. Our calculations thus suggest that the \sone{} and \ctone{} excitons are of Frenkel-like nature. Their energy difference originates from different excitation symmetries possible for the H$\rightarrow$L transition (namely $t_{1g}$, $t_{2g}$, and $g_g$ for the \sone{} and $h_g$ for the \ctone{}).\cite{kobayashi_wannier-like_2020} 

In contrast to the \sone{} and \ctone{} excitons, the momentum map of the \cttwo{} band shows a much more star-shaped POT fingerprint in both theory and experiment (Fig.~\ref{fig:comparison}c,f). This is to be expected, since the electronic part of the \cttwo{} excitons contains not only contributions from the LUMO orbital, but also contributions from the LUMO+1 orbital. Note, however, that we find the experimentally observed star-shaped pattern to be only partially reproduced by the $GW$+BSE calculation. An indication towards the cause of this discrepancy is found by considering the electron-hole separation of the excitons making up the \cttwo{} band. Here, we find that the positions of the electron and the hole contributions are strongly anticorrelated (Fig.~\ref{fig:comparison}i), with the electron confined to the neighboring molecule of the dimer. In fact, the mean electron-hole separation is as large as 7.6~\AA, which is close to the core-to-core distance of the \csixty{} molecules. Although these theoretical results confirm the previously-reported charge-transfer nature of the \cttwo{} excitons \cite{stadtmuller_strong_2019, emmerich_ultrafast_2020}, they also reflect the limitations of the \csixty{} dimer approach. Indeed, the dimer represents the minimal model to account for an intermolecular exciton delocalization effect, but it cannot fully account for dispersion effects\cite{haag_signatures_2020} (cf. Extended Fig. 5), which are required for a quantitative comparison with experimental data. Besides the discrepancy in the \cttwo{} momentum map, this could also be an explanation why the \ctone{} in the present work is of Frenkel-like nature, but could have charge-transfer character according to previous studies\cite{stadtmuller_strong_2019, emmerich_ultrafast_2020}.
However, future developments will certainly allow scaling up of the cluster size in the calculation, so that exciton wavefunctions with larger electron-hole separation can be accurately described.
Most importantly, we find that the present dimer $GW$+BSE calculations are clearly suited to elucidate the multiorbital character of the excitons, which is an indispensable prerequisite for the correct interpretation of tr-POT data of excitons in organic semiconductors. 

\section{CONCLUSION}
In conclusion, we have shown how the energy- and momentum-resolved photoemission spectrum of excitons in an organic semiconductor depends on the multiorbital nature of these excitons. By extending POT to fully-interacting exciton states calculated in the framework of the Bethe-Salpeter equation, we found that the energy of the photoemitted electron of the exciton quasiparticle is determined by the position of the hole and the exciton energy in combination with the probe photon energy. This leads to the prediction of multiple peaks in the photoelectron spectrum, which we verify experimentally, and allows disentangling the different orbital contributions, the wavefunction localization, and the charge-transfer character. Similarly, the momentum fingerprint provides access to the electron states that make up the exciton. Most importantly, we introduce time-resolved photoemission orbital tomography as a key technique for the study of exciton wavefunctions in organic semiconductors.

\section{Acknowledgements}
This work was funded by the Deutsche Forschungsgemeinschaft (DFG, German Research Foundation) - 432680300/SFB 1456, project B01 and 217133147/SFB 1073, projects B07 and B10. G.S.M.J. acknowledges financial support by the Alexander von Humboldt Foundation. 
A.W., C.S.K., and P.P acknowledge support from the Austrian Science Fund (FWF) project I 4145. The computational results presented were achieved using the Vienna Scientific Cluster (VSC) and the local high-performance resources of the University of Graz.
R.H., M.A., and B.S. acknowledge financial support by the DFG - 268565370/TRR 173, projects B05 and A02. B.S. acknowledges further support by the Dynamics and Topology Center funded by the State of Rhineland-Palatinate.

\section{Author Contributions}
D.St., M.R., S.S., M.A., B.S., P.P.,  G.S.M.J. and S.M. conceived the research. W.B., D.Sch. and J.P.B. carried out the time-resolved momentum microscopy experiments. W.B. analyzed the data. W.B. and R.H. prepared the samples. A.W., C.S.K., G.D.A, X.B. and P.P. performed the calculations and analyzed the theoretical results. All authors discussed the results. G.S.M.J. and S.M. were responsible for the overall project direction and wrote the manuscript with contributions from all co-authors. 

\bibliography{2022c60exciton} 

\newpage
\section{Methods}

\subsection{Femtosecond momentum microscopy of \csixty{}/Cu(111)}
\begin{figure}[hp]
    \centering
    \includegraphics[width=\linewidth]{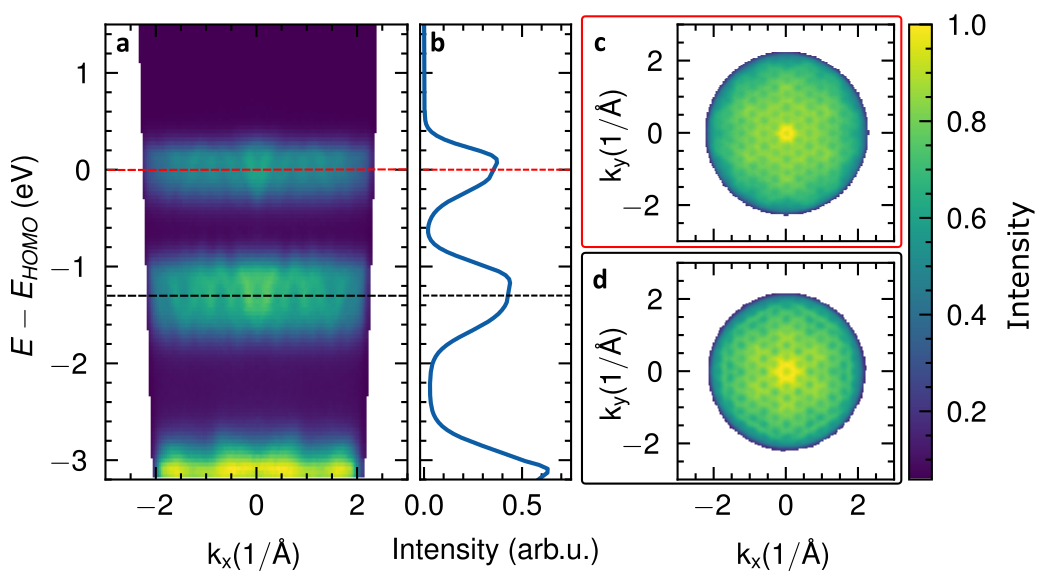}
    \caption{Static photoelectron momentum microscopy of the multilayer \csixty{} sample, taken from the time-resolved data at -1000~fs. The high contrast and line shape of the occupied molecular orbitals as shown in the energy-momentum cut (\textbf{a}) and integrated energy distribution curve (\textbf{b}) confirm that an influence of the Cu(111) substrate can be ignored. \textbf{c, d:} Furthermore, the clear modulation of the HOMO (\textbf{c}) and HOMO-1 (\textbf{d}) momentum maps confirm the high crystallinity of the multilayer at 80~K\cite{haag_signatures_2020}}
    \label{fig:occupied_states}
\end{figure}

We apply full multidimensional time- and angle-resolved photoelectron spectroscopy (tr-ARPES) to a multilayer \csixty{} crystal evaporated onto Cu(111), where the film thickness was such that no photoemission signature of the underlying Cu(111) could be observed in our experiment. We verified the sample quality by performing momentum microscopy of the occupied HOMO and HOMO-1 states simultaneously to the measurement of the excited states (see Extended Fig.~\ref{fig:occupied_states}). Femtosecond exciton dynamics were induced using $\approx$100~fs, $h\nu =$~2.9~eV or $\approx$100~fs, $h\nu =$~3.6~eV laser pulses derived from the frequency-doubled output of a optical parametric amplifier. The exciton dynamics were probed using our custom photoemission momentum microscope with a 500~kHz ultrafast 26.5~eV extreme ultraviolet (EUV) light source \cite{keunecke_time-resolved_2020} that enables us to map the photoelectron momentum distribution over the full photoemission horizon in a kinetic energy range exceeding 6~eV and an overall time resolution of $\approx$100~fs (the EUV pulse length is about 20~fs). The pump fluence was set to $90(10)$~$\mu$J/cm$^2$ and $20(5)$~$\mu$J/cm$^2$ for the $h\nu =$~2.9~eV and the $h\nu =$~3.6~eV measurement, respectively.
To prevent the free rotation of \csixty{} molecules, we cooled the sample down to $\approx$80~K \cite{wang_orientational_2001}. In addition to the resulting long-range periodic ordering of the \csixty{} crystal, cooling was also observed to prevent light-induced polymerization.  

\subsection{Momentum microscopy data analysis}
\subsubsection{Space-charge effects}
In the momentum microscopy experiment, a balance has to be found between sufficiently low pump and probe light intensities to avoid space-charge effects, but also having sufficient intensity for the optical excitation (pump) and reasonably short integration times (probe). Most of the time, for our settings, small space-charge effects are present in the data, but do not lead to strong distortions in the band-structure data and can be easily corrected. Therefore, the first step in the data analysis was to subtract a space-charge-induced delay- and momentum-dependent kinetic energy shift. For this purpose, the central kinetic energy of the HOMO was determined for normalization. To avoid the influence of the \csixty{} crystal band structure \cite{haag_signatures_2020}, we fitted a two-dimensional Lorentzian \cite{schonhense_multidimensional_2018} and shifted the kinetic energy distribution accordingly, leading to the expected overall flat shape of the molecular orbitals in the ARPES data. 

\subsubsection{Replicas from the 13\textsuperscript{th} harmonic}
Although we use narrow-band multilayer mirrors to select the 11\textsuperscript{th} harmonic at $h\nu$ = 26.5~eV from our laser-based high-harmonic generation spectrum\cite{keunecke_time-resolved_2020}, we observe subtle replicas of the HOMO, HOMO-1, and HOMO-2 states in the unoccupied regime of the spectrum that are caused by photoemission from the 13\textsuperscript{th} harmonic at at $h\nu$ = 31.2~eV. 
To quantify these replica signals, we fitted the static reference spectrum above $E-E_H = 1.2$~eV (i.e., in the unoccupied regime of the spectrum) with three Gaussian-shaped peaks for the HOMO replicas and an exponential function to account for residual photoemission intensity in the unoccupied regime that is caused in this spectral region by the much stronger direct 11\textsuperscript{th} harmonic one-photon-photoemission from the HOMO state. After carrying out this fitting routine, we are able to calculate clean 11\textsuperscript{th} harmonic spectra (static and time-resolved) via subtraction of the fitted 13\textsuperscript{th} harmonic HOMO replicas. Note that we only subtract the replica signals, but not the background signal that is caused by one-photon-photoemission with the 11\textsuperscript{th} harmonic from the HOMO state, because this background is time-dependent\cite{stadtmuller_strong_2019}, and needs to be explicitly considered in the fitting procedure. The data shown in Fig.~3 of the main text is processed in the way described above. 

\subsubsection{Fitting procedure for the time-resolved data}
From replica-free trPES data for $h\nu =$~2.9~eV excitation, we determine the amplitude $A_i$, kinetic energy $E_i$, and bandwidth $\Delta E_i$ for the $i$\textsuperscript{th} exciton signature using a global fitting approach. In particular, we apply the model
\begin{equation}
\begin{split}
    I(E,t) & = \frac{A_{\cttwo{}}(t)}{\sqrt{2\pi}\Delta E_{\cttwo{}}} \mathrm{exp}\left[(E-E_{\cttwo{}}(t))^2/\Delta E_{\cttwo{}}^2\right] \\
     & + \frac{A_{\ctone{}}(t)}{\sqrt{2\pi}\Delta E_{\ctone{}}} \mathrm{exp}\left[(E-E_{\ctone{}})^2/\Delta E_{\ctone{}}^2\right]    \\
     & + \frac{A_{\sone{}}(t)}{\sqrt{2\pi}\Delta E_{\sone{}}} \mathrm{exp}\left[(E-E_{\sone{}})^2/\Delta E_{\sone{}}^2\right] \\ 
     & + A_{bg}(t)*\mathrm{exp}\left[-E/\tau\right].
\end{split}
\label{eq:fitmodel_trPES}
\end{equation}
Here, the last term is needed to account for the above-mentioned delay-dependent photoemission intensity that is caused by a transient renormalization of the HOMO state, as found in Ref.~\onlinecite{stadtmuller_strong_2019}.  

\begin{figure}[hp]
    \centering
    \includegraphics[width = \linewidth]{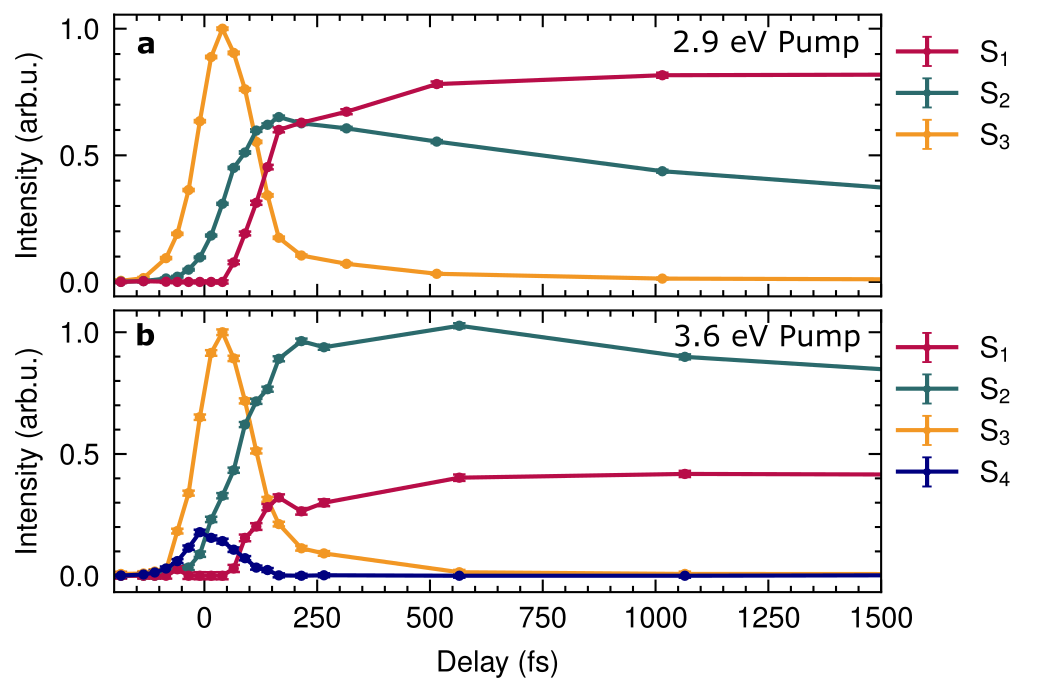}
    \caption{\textbf{a, b:} Exciton-resolved measurement of the femtosecond relaxation dynamics after $h\nu =$2.9 and $h\nu =$3.6~eV excitation, respectively. The relative intensities were acquired by fitting the model in Eq.~\ref{eq:fitmodel_trPES} and \ref{eq:fitmodel_trPES_extension} to the momentum-integrated trPES data. Analysis of the time dependence indicates a sub 50~fs lifetime of the \stwo{} states. The comparably large \ctone{} population in the $h\nu =$~3.6~eV measurement can potentially be explained by a relaxation of the \stwo{} into both the \cttwo{} and \ctone{} states.}
    \label{fig:expdynamics}
\end{figure}

The fit results of this model applied to the $h\nu =$~2.9~eV excitation and momentum-integrated data are shown in Table~\ref{tab:fit_trPES}, and Extended Fig.~\ref{fig:expdynamics}a for the time-resolved exciton dynamics.

\begin{table}[h!]
\centering
\begin{tabular}{|c | c | c |} 
\hline
Exciton & Kin. Energy (eV) & Bandwidth (FWHM) (eV) \\ [0.5ex] 
\hline\hline
\cttwo{} (t=0~fs) & 2.768(2) & 0.606(4) \\ 
\ctone{} & 1.978(2) & 0.406(3) \\
\sone{} & 1.667(1) & 0.362(2) \\
\hline
\end{tabular}
\caption{ 
\label{tab:fit_trPES}
Peak parameters for $h\nu =$~2.9~eV excitation, extracted using Eq.~\ref{eq:fitmodel_trPES}. Note that we give the full width at half maximum (FWHM) for the bandwidth. }
\end{table}

For the measurement with $h\nu =$~3.6~eV excitation, we account for the \stwo{} exciton band by extending the model in Eq.~\ref{eq:fitmodel_trPES} with a set of Gaussian peaks with identical temporal evolution, given by
\begin{equation}
\begin{split}
    I(E,t) = ... + \frac{A_{\stwo{}}(t)}{2\sqrt{2\pi}\Delta E_{\stwo{}}} ( &\mathrm{exp}\left[(E-E_{\stwo{},upper})^2/\Delta E_{\stwo{}}^2\right] \\
    +& \mathrm{exp}\left[(E-E_{\stwo{},lower})^2/\Delta E_{\stwo{}}^2\right]),
\end{split}
\label{eq:fitmodel_trPES_extension}
\end{equation}
which follows the same notation as Eq.~\ref{eq:fitmodel_trPES}. Here, we set $E_{\stwo{},upper}$ to be close to 3.6~eV, and following the $GW$+BSE calculation we set $E_{\stwo{},upper} - E_{\stwo{},lower} = 1.4$~eV. 
Fitting this model to the momentum-integrated $h\nu =$~3.6~eV excitation data, we find $E_{\stwo{},upper} = 3.59(1)$~eV, and for the FWHM of the \stwo{} we find $0.58(3)$~eV. The time-resolved amplitudes retrieved using this model are shown in Extended Fig.~\ref{fig:expdynamics}b. Furthermore, this analysis was used in Fig.~3 of the main text to subtract the exponential background $A_{bg}(t)*\mathrm{exp}\left[-E/\tau\right]$ related to the transient broadening of the HOMO state. 

\subsubsection{Fitting procedure for the time- and momentum-resolved data}
\begin{figure}[hp]
    \centering
    \includegraphics[width = \linewidth]{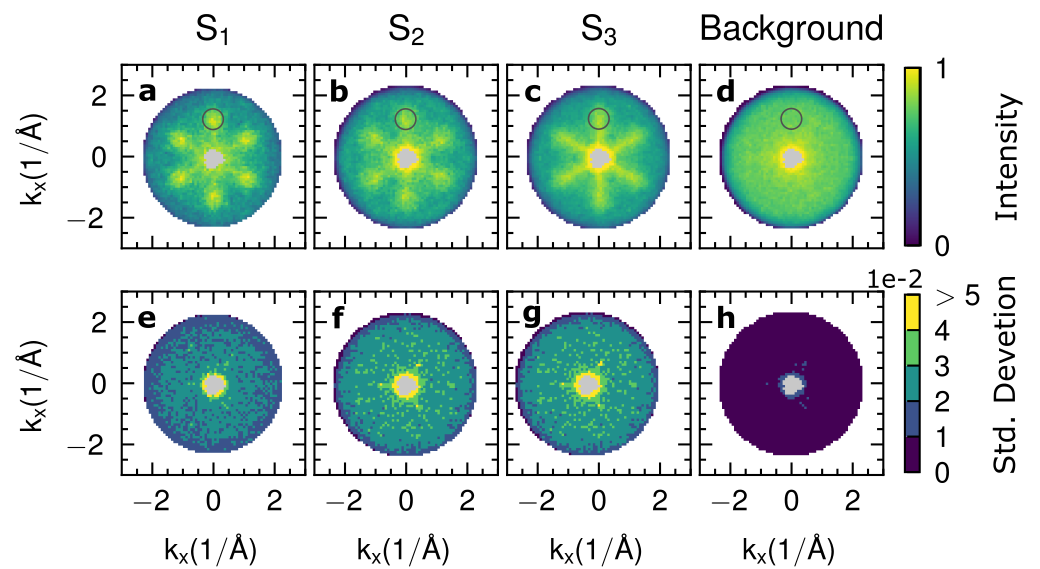}
    \caption{Amplitude (\textbf{a,b,c,d}) and standard deviation (\textbf{e,f,g,h}) of the extracted momentum maps of the \sone{}, \ctone{}, \cttwo{} and the exponential background ($A_{bg}(k)$), respectively. The data are scaled to the mean value inside the circled area.}
    \label{fig:kmap_errors}
\end{figure}

In order to analyze the time-resolved data also momentum-resolved and thereby retrieve the momentum patterns that are shown in Fig.~\ref{fig:exp_excitons} and Fig.~\ref{fig:comparison} in the main text, we carry out the fitting routine separately for pixel-resolved energy-distribution curves 
in the momentum distribution (1 pixel corresponds to $\approx$ 0.02~\AA\textsuperscript{-2}). To optimize the signal-to-noise ratio, we apply a three-fold rotational symmetrization and a mirror symmetrization that match with crystal symmetry. Nevertheless, the signal-to-background ratio in a small region around the center of the photoemission horizon remains below 1, and we therefore exclude this region in our analysis (grey areas in Fig.~4 in the main text). Also, for the momentum-resolved data, the replica HOMO background signals due to the 13\textsuperscript{th} harmonic amounts to 0, 1 or at most 2 counts in the pixel-resolved (momentum-resolved) energy-distribution curves, and can therefore not be fitted and subtracted accurately as described above for the momentum-integrated data. As such, we need to ignore the HOMO replicas from the 13\textsuperscript{th} harmonic in the momentum-resolved analysis. 
We avoid overfitting of the model in Eq.~\ref{eq:fitmodel_trPES} by fixing the energy and bandwidth of the peaks in the fitting routine to the parameters given in Table~\ref{tab:fit_trPES}. 
Thus, the set of free parameters in the momentum-resolved fitting procedure is limited to $A_{\cttwo{}}(k)$, $A_{\ctone{}}(k)$, $A_{\sone{}}(k)$ and $A_{bg}(k)$. This approach enables the extraction of reliable momentum distributions also for the partially overlapping energy distributions of the \sone{} and \ctone{}. The $1\sigma$ errors for the full momentum maps are shown in Extended Fig.~\ref{fig:kmap_errors}. 

We note that the overall photoemission intensity of the \stwo{} peak in the $h\nu =$~3.6~eV excitation data is comparably low due to the sub-50~fs decay to the lower-energy \cttwo{} excitons (see Fig.~3 in main text, Extended Fig.~6b). Furthermore, with $h\nu =$~3.6~eV excitation, two-photon photoemission with 2~$*$~3.6~eV = 7.2~eV is sufficient to overcome the work function, so that space-charge effects could only be avoided by considerably reducing the $h\nu =$~3.6~eV pump intensity. Therefore, the signal-to-noise ratio in these measurements was not sufficient for a momentum-resolved analysis of the \stwo{} exciton data.

\subsection{Calculation of the \csixty{} exciton spectrum}
The \emph{ab initio} calculation of the exciton spectrum of the \csixty{} film was performed in two steps, using a $GW$+BSE approach.
For the static electronic structure, we perform calculations for two unique \csixty{} dimers, which have been extracted from the known structure of the molecular film \cite{david_structure_1991, david_structure_1992} (see Fig.~\ref{fig:bse_excitons}c, dimers 1-2 and 1-4 respectively). Starting from Kohn-Sham orbitals and energies of a ground state DFT calculation (6-311G*/PBE0+D3) \cite{krishnan_basisset_1980, frisch_basisset_1984, adamo_pbe0_1999, grimme_d3_2010} using ORCA 5.0.1 \cite{neese_orca_2012, neese_update5_2022}, we employ the Fiesta code \cite{jacquemin_is_2017} to self-consistently correct the molecular energy levels by quasi-particle self-energy calculations with the $GW$ approximation. To account for polarization effects beyond the molecular dimer, we embed the dimer cluster in a discrete polarizable model using the MESCal program \cite{davino_mescal_2014, li_bse_2016, li_correlated_2017}. We found that mimicking 2 layers of the surrounding \csixty{} film in such a way resulted in the convergence of the band gap within 0.1~eV with a removal energy from the highest valence level of 6.65~eV. The close agreement of this quasi-particle energy with the experimentally determined work function of 6.5~eV gives us additional confidence in the choice of our embedding environment. The calculated quasi-particle energy levels are shown in Fig.~\ref{fig:bse_excitons}a. Here the finite width of the black bars actually arises from multiple energy levels forming bands on the energy axis. We characterize them according to symmetry \cite{dresselhaus_c60_1996} as HOMO-1, HOMO, LUMO, and LUMO+1 bands, each consisting of 18, 10, 6 and 6 energy levels per dimer, respectively. Note that we combine HOMO-1 is made up by states from two different irreducible representations of the isolated gas phase \csixty{} molecule which are practically forming a single band.

Building upon the $GW$ energies, we compute neutral electron-hole excitations by solving the Bethe-Salpeter equation beyond the Tamm-Dancoff approximation (TDA). This yields the excitation energies $\Omega_m$ and the electron-hole coupling coefficients $X^{(m)}_{vc}$ for a series of excitons labelled with $m$. 
We first analyze the resulting optical absorption spectrum which is shown in Fig.~\ref{fig:bse_excitons}c as black solid line. It reveals a prominent absorption band around $h\nu =$~3.6~eV that is well-known from gas-phase spectroscopy\cite{dai_ultravioletvisible_1994}. Secondly, the dimer calculation reveals a strong optical absorption at $h\nu =$~2.8~eV as well as a weakly dipole-allowed transition at $h\nu =$~2~eV. Both of these transitions are known to only appear in aggregated phases of \csixty{} \cite{wang_aggregates_1993}, and cannot be understood by considering only a single \csixty{} molecule. 
We refer to the exciton bands around $\Omega =$~1.9, 2.1, 2.8 and 3.6~eV as \sone{}, \ctone{}, \cttwo{} and \stwo{}, respectively. 

In line with our classification of the $GW$ energy levels, one can group the composition of the excitons into four categories according to the contributing quasi-particle energy levels. As visualized in Fig.~\ref{fig:bse_excitons}c, this shows that \sone{} and \ctone{} are almost completely described by HOMO to LUMO transitions. On the other hand, \cttwo{} is predicted to have a small contribution of HOMO-1 character, however this contribution is too small to be reliably measured in our experiment. Finally, for \stwo{} we observe a clear and almost equal mixture of HOMO-1 and HOMO contributions, which is also confirmed by our measurements. Contributions from lower lying valence bands are negligibly small in the studied energy window.

\subsection{Calculation of the exciton momentum maps}
Based on our Kohn-Sham orbitals and BSE excitation coefficients, we calculate theoretical momentum maps for each exciton according to Eq.~\ref{eq:trPOT} following the derivation of Kern \textit{et al.}\cite{kern_exciton_tomog}. Note that for better readability, Eqs.~\eqref{eq:exciton_state} and \eqref{eq:trPOT} are given within the TDA, however, can in general be extended to include also de-excitation terms.\cite{kern_exciton_tomog} In the present case, we found the de-excitation contributions to be marginal (below 1\%) without affecting the appearance of the momentum maps and our interpretation. The energy conservation term comprises the BSE excitation energies ($\Omega_m$), the $GW$ quasi-particle energies for electron removal (i.e. ionization potential $\varepsilon_v$), and the probe energy of $h\nu =$~26.5~eV in accordance with our experimental setup. Furthermore, we include an inner potential to correct for the photoemission intensity variation of 3D molecules along the moment vector component perpendicular to the surface. Here, we choose a value of 12.5~eV, which has already been shown to match with experimental \csixty{} data \cite{hasegawa_potential_1998, haag_signatures_2020}. (Note that while considering the inner potential of the film is essential to describe the ARPES fingerprint, a variation of the inner potential between 12 and 14~eV indicated no influence on the interpretation of our results.)
As we exploited the plane-wave approximation, the calculated photoemission intensity is modulated by the momentum-dependent polarization factor $\left| \ve{A} \cdot \ve{k} \right|$, which we modelled as p-polarized light incoming with 68$^{\circ}$ to the surface normal according to experiments. To account for the symmetry of the \csixty{} film, the momentum maps were 3-fold rotated and mirrored. Finally, application of Eq.~\ref{eq:trPOT} provides us with a 4D data set of simulated photoemission intensity as a function of the excitation energy ($\Omega_m$), the kinetic energy ($E_{kin}$) and the momentum components $k_x$ and $k_y$. Analogous to experiment, we referenced the kinetic energy against the energy of the HOMO. The calculated ionization potential was further used to set the photoemission horizon of the theoretical momentum maps. Finally, to arrive at the theoretical momentum maps shown in Fig. \ref{fig:comparison}, we sum up the photoelectron intensities of each contributing electron-hole transition in an kinetic energy range of 200~meV centered on the respective exciton band.

\section{Extended Figures}

\end{document}